\documentclass[conference]{IEEEtran}
\usepackage{cite}
\usepackage{amsmath,amssymb,amsfonts}
\usepackage{algorithmic}
\usepackage{graphicx}
\usepackage{textcomp}
\usepackage{xcolor}
\usepackage[braket, qm]{qcircuit} 

\usepackage[utf8]{inputenc}

\begin{document}

\title{The PPKN Gate: An Optimal 1-Toffoli Input-Preserving Full Adder for Quantum Arithmetic}

\author{\IEEEauthorblockN{G. Papakonstantinou}
\IEEEauthorblockA{\textit{School of Electrical and Computer Engineering} \\
\textit{National Technical University of Athens}\\
Athens, Greece \\
papakon@cs.ntua.gr}
}

\maketitle

\begin{abstract}
Efficient arithmetic operations are a prerequisite for practical quantum computing. Optimization efforts focus on two primary metrics: Quantum Cost (QC), determined by the number of non-linear gates, and Logical Depth, which defines the execution speed. Existing literature identifies the HNG gate as the standard for Input-Preserving Reversible Full Adders. HNG gate typically requires a QC of 12 and a logical depth of 5, in the area of classical reversible circuits. This paper proposes the \textbf{PPKN Gate}, a novel design that achieves the same input-preserving functionality using only \textbf{one Toffoli gate} and five CNOT gates. With a Quantum Cost of 10 and a reduced \textbf{logical depth of 4}, the PPKN gate outperforms the standard HNG gate in both complexity and speed. Furthermore, we present a modular architecture for constructing an $n$-bit Ripple Carry Adder by cascading PPKN modules.
\end{abstract}


\section{Introduction}
In reversible logic synthesis, the Full Adder is the cornerstone of complex arithmetic units. Since the foundational work by Toffoli \cite{Toffoli80}, minimizing the number of non-linear gates (Toffoli/Fredkin) has been the primary objective, as these gates are significantly more expensive to implement in fault-tolerant quantum technologies than linear gates (CNOTs) \cite{NielsenChuang}.

A specific class of adders, known as Input-Preserving Full Adders, maps inputs $(C_{in}, A, B, 0)$ to $(Sum, A, B, C_{out})$. This preservation of operands is essential in quantum algorithms to avoid uncomputation penalties. The state-of-the-art design for this specification has long been the \textbf{HNG Gate} proposed by Haghparast and Navi \cite{HNG} in quantum and classical area. While efficient, the HNG gate relies on two Toffoli gates (QC=12) and requires a logical depth of 5 to manage operand preservation.

This work introduces the \textbf{PPKN Gate} using only cnot and Toffoli gates. By optimizing the linear transformation stages, the PPKN gate utilizes a single Toffoli gate to compute the Carry. This topology not only reduces the Quantum Cost to 10 but also parallelizes linear operations to achieve a logical depth of 4.

\section{The Proposed PPKN Gate}

The PPKN gate is a $4 \times 4$ reversible circuit designed to minimize quantum cost, gate count, and logical depth.

\subsection{Circuit Specification}
\begin{itemize}
    \item \textbf{Inputs:} $L_1 = C_{in}$, $L_2 = A$, $L_3 = B$, $L_4 = 0$ (Ancilla).
    \item \textbf{Outputs:} $L_1 = Sum$, $L_2 = A$, $L_3 = B$, $L_4 = C_{out}$.
\end{itemize}

\subsection{Gate Netlist and Depth Analysis}
The circuit consists of 6 gates. The logical depth is determined by organizing operations into parallel time steps (T). Gates in the same time step operate on disjoint quantum lines (Fan-out from a shared control is allowed).

\begin{enumerate}
    \item \textbf{Time Step 1 (Depth 1):} 
    \begin{itemize}
        \item \textbf{Gate 1:} CNOT($L_3, L_1$). Target: $C_{in}$.
        \item \textbf{Gate 2:} CNOT($L_3, L_2$). Target: $A$.
    \end{itemize}
    \textit{Parallel Execution: Shared Control $L_3$, distinct targets.}
    
    \item \textbf{Time Step 2 (Depth 2):} 
    \begin{itemize}
        \item \textbf{Gate 3:} Toffoli($L_1, L_2, L_4$). Target: $C_{out}$.
    \end{itemize}
    
    \item \textbf{Time Step 3 (Depth 3):} 
    \begin{itemize}
        \item \textbf{Gate 4:} CNOT($L_3, L_2$). Target: $A$ (Restoration).
        \item \textbf{Gate 5:} CNOT($L_3, L_4$). Target: $C_{out}$ (Correction).
    \end{itemize}
    \textit{Parallel Execution: Shared Control $L_3$, distinct targets.}

    \item \textbf{Time Step 4 (Depth 4):} 
    \begin{itemize}
        \item \textbf{Gate 6:} CNOT($L_2, L_1$). Target: $Sum$.
    \end{itemize}
    \textit{Dependency: Must wait for $L_2$ restoration in Step 3.}
\end{enumerate}

\textbf{Total Logical Depth: 4.}

\subsection{Circuit Diagram}
\begin{figure}[htbp]
    \centering
    \[
    \Qcircuit @C=1.0em @R=1.5em {
    \lstick{L_1: C_{in}} & \targ     & \qw       & \ctrl{3} & \qw       & \qw      & \targ     & \rstick{Sum} \qw \\
    \lstick{L_2: A}      & \qw       & \targ     & \ctrl{2} & \targ     & \qw      & \ctrl{-1} & \rstick{A} \qw \\
    \lstick{L_3: B}      & \ctrl{-2} & \ctrl{-1} & \qw      & \ctrl{-1} & \ctrl{1} & \qw       & \rstick{B} \qw \\
    \lstick{L_4: 0}      & \qw       & \qw       & \targ    & \qw       & \targ    & \qw       & \rstick{C_{out}} \qw
    }
    \]
    \caption{The Proposed PPKN Gate Structure.}
    \label{fig:ppkn_circuit}
\end{figure}
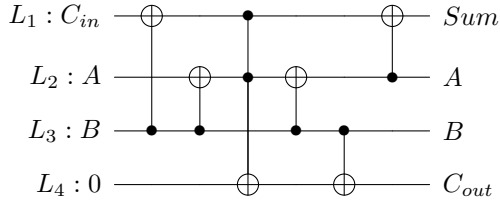

The quantum circuit diagram is illustrated in Fig. \ref{fig:ppkn_circuit}.

\subsection{Logic Verification}
We verify the functionality by tracing the boolean states.

\subsubsection{Step 1 (Pre-computation)}
\begin{align*}
L_1 & \leftarrow C_{in} \oplus B \\
L_2 & \leftarrow A \oplus B
\end{align*}

\subsubsection{Step 2 (Non-Linear Interaction)}
The Toffoli gate uses the transformed lines 1 and 2 to toggle line 4.
\[ L_4 = 0 \oplus (L_1 \cdot L_2) = (C_{in} \oplus B)(A \oplus B) \]
Expanding this term:
\[ L_4 = C_{in}A \oplus C_{in}B \oplus AB \oplus B \]

\subsubsection{Step 3 (Restoration and Correction)}
We restore $L_2$ ($A$) and correct $L_4$ using Control $L_3$ ($B$):
\begin{align*}
L_2 & \leftarrow (A \oplus B) \oplus B = \mathbf{A} \\
L_4 & \leftarrow (C_{in}A \oplus C_{in}B \oplus AB \oplus B) \oplus B
\end{align*}
The linear $B$ terms on $L_4$ cancel out:
\[ L_4 = C_{in}A \oplus C_{in}B \oplus AB = \mathbf{C_{out}} \]

\subsubsection{Step 4 (Sum Generation)}
Finally, we use the restored $A$ (on $L_2$) to target $L_1$:
\[ L_1 \leftarrow (C_{in} \oplus B) \oplus A = \mathbf{Sum} \]

The final state is $\{Sum, A, B, C_{out}\}$.

\section{Performance Comparison}

We compare the PPKN gate against the standard HNG gate \cite{HNG} and the TSG gate \cite{TSG}. The results are summarized in Table \ref{tab:comparison}. We note that the comparison is based using only not, cnot and Toffoli gates (NCT library).

\begin{table}[htbp]
\caption{Comparison: PPKN vs Existing Gates}
\begin{center}
\begin{tabular}{|c|c|c|c|}
\hline
\textbf{Metric} & \textbf{HNG Gate} \cite{HNG} & \textbf{TSG Gate} \cite{TSG} & \textbf{PPKN Gate} \\
\hline
Inputs Preserved & Yes & Yes & \textbf{Yes} \\
\hline
Total Gate Count & 5 & 6 & \textbf{6} \\
\hline
\textbf{Toffoli Count} & 2 & 2 & \textbf{1} \\
\hline
\textbf{Quantum Cost} & 12 & 14 & \textbf{10} \\
\hline
\textbf{Logical Depth} & 5 & 6 & \textbf{4} \\
\hline
\end{tabular}
\end{center}
\label{tab:comparison}
\end{table}

\textbf{Analysis:} The PPKN gate outperforms the HNG gate in both primary metrics. The HNG gate requires a logical depth of 5 because the $C_{in}$ line acts as a Control for the carry calculation and a Target for the sum calculation, forcing those operations to be sequential. The PPKN gate avoids this conflict, allowing for a depth of 4. Furthermore, the PPKN gate reduces the Quantum Cost by 17\% (10 vs 12).

\section{Scalability: N-Bit Ripple Carry Adder}

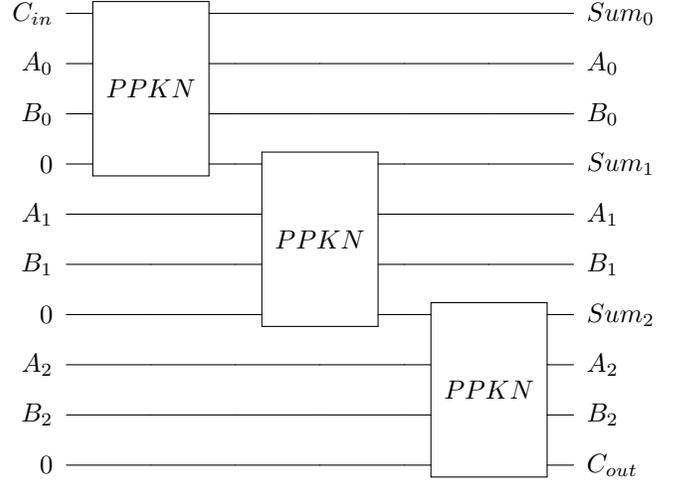
\begin{figure}[htbp]
    \centering
    \[
    \Qcircuit @C=1.0em @R=1.0em {
    \lstick{C_{in}} & \multigate{3}{PPKN} & \qw & \qw & \qw & \qw & \rstick{Sum_0} \qw \\
    \lstick{A_0}    & \ghost{PPKN}        & \qw & \qw & \qw & \qw & \rstick{A_0} \qw \\
    \lstick{B_0}    & \ghost{PPKN}        & \qw & \qw & \qw & \qw & \rstick{B_0} \qw \\
    \lstick{0}      & \ghost{PPKN}        & \qw & \multigate{3}{PPKN} & \qw & \qw & \rstick{Sum_1} \qw \\
    \lstick{A_1}    & \qw                 & \qw & \ghost{PPKN}        & \qw & \qw & \rstick{A_1} \qw \\
    \lstick{B_1}    & \qw                 & \qw & \ghost{PPKN}        & \qw & \qw & \rstick{B_1} \qw \\
    \lstick{0}      & \qw                 & \qw & \ghost{PPKN}        & \qw & \multigate{3}{PPKN} & \rstick{Sum_2} \qw \\
    \lstick{A_2}    & \qw                 & \qw & \qw                 & \qw & \ghost{PPKN}        & \rstick{A_2} \qw \\
    \lstick{B_2}    & \qw                 & \qw & \qw                 & \qw & \ghost{PPKN}        & \rstick{B_2} \qw \\
    \lstick{0}      & \qw                 & \qw & \qw                 & \qw & \ghost{PPKN}        & \rstick{C_{out}} \qw
    }
    \]
    \caption{3-bit Ripple Carry Adder using cascaded PPKN Gates. The internal carry signals propagate from the bottom of one stage to the top of the next. All outputs are preserved and available at the final stage.}
    \label{fig:rca}
\end{figure}

The topology of the PPKN gate allows for efficient cascading. Figure \ref{fig:rca} demonstrates a 3-bit Ripple Carry Adder. The Carry-Out ($L_4$) of block $i$ connects to the Carry-In ($L_1$) of block $i+1$, while inputs $A$ and $B$ are preserved.

\section{Conclusion}
This paper presented the PPKN Gate, an optimal 1-Toffoli Reversible Full Adder. By leveraging an improved linear cancellation technique, the PPKN gate breaks the 2-Toffoli barrier for input-preserving adders. It achieves a Quantum Cost of 10 and a Logical Depth of 4, proving explicitly superior to the standard HNG gate for high-performance quantum arithmetic.

\end{document}